\documentclass[prb,twocolumn,preprintnumbers,amsmath,amssymb]{revtex4}
\usepackage{graphicx}
\usepackage{dcolumn}
\usepackage{bm}
\usepackage{epstopdf}
\usepackage{amsmath}
\usepackage[dvips]{color}

\begin{document}

\title{Magnetic properties, spin waves and interaction between spin excitations and
2D electrons in interface layer in Y$_3$Fe$_5$O$_{12}$ / AlO${ }_{x}$ / GaAs-heterostructures }
\author{L.V. Lutsev$^{1}$}
\author{A.I. Stognij$^{2}$}
\author{N.N. Novitskii$^{2}$}
\author{V.E. Bursian$^{1}$}
\author{A. Maziewski$^{3}$}
\author{R. Gieniusz$^{3}$}

\affiliation{$^{1}$Ioffe Physical-Technical Institute, 194021, St. Petersburg, Russia}
\affiliation{$^{2}$Scientific and Practical Materials Research
Centre, National Academy of Sciences of Belarus, 220072, Minsk,
Belarus}
\affiliation{$^{3}$Faculty of Physics, University of
Bialystok, 15-097, Bialystok, Poland}
\date{\today}

\begin{abstract}
We describe synthesis of submicron Y${ }_3$Fe${ }_5$O${ }_{12}$
(YIG) films sputtered on GaAs-based substrates and present results
of the investigation of ferromagnetic resonance (FMR), spin wave
propagation and interaction between spin excitations and 2D
electrons in interface layer in YIG / AlO${ }_{x}$ /
GaAs-heterostructures. It is found that the contribution of the
relaxation process to the FMR linewidth is about 2~\% of the
linewidth $\Delta H$. At the same time, for all samples FMR
linewidths are high. Sputtered YIG films have magnetic
inhomogeneity, which gives the main contribution to the FMR
linewidth. Transistor structures with two-dimensional electron gas
(2DEG) channels in AlO${ }_{x}$ / GaAs interface governed by
YIG-film spin excitations are designed. An effective influence of
spin excitations on the current flowing through the GaAs 2DEG
channel is observed. Light illumination results in essential changes
in the FMR spectrum. It is found that an increase of the 2DEG
current leads to an inverse effect, which represents essential
changes in the FMR spectrum.

\end{abstract}

\maketitle

\email{l_lutsev@mail.ru}

\section{Introduction}

Integration of ferrites with semiconductors offers many advantages
and new possibilities in microwave applications such as high-speed
wireless communications, active phased array antennas for radars,
astronomy systems, auto radars, space electronics and satellite
navigation. This integration gives significant advantages in
miniaturisation, bandwidth, speed, radio reception selectivity and
the production costs of monolithic microwave integrated circuits
(MMICs)~\cite{Chen12}. Ferrite film growth on semiconductor
substrates is very important for development of new types of
spintronic and spin-wave devices such as microwave filters, delay
lines, and spin-polarized field-effect transistors (spin-FET).

At present, spin-wave devices have been realized on the basis of Y${
}_3$Fe${ }_5$O${ }_{12}$ (YIG) films grown on gadolinium-gallium
garnet (Gd${ }_3$Ga${ }_5$O${ }_{12}$, GGG)
substrates~\cite{Stan09,Kabos}. Narrow-band filtration can be
achieved in YIG-based one-dimensional magnonic
crystals~\cite{Mrucz14,Bes15,Vys17}. The pulsed laser deposition technique has
been used to grow submicron YIG films on GGG substrates for
microwave spin-wave band pass filter~\cite{Man09}. Construction of
spin-wave devices on the basis of YIG films directly deposited on
semiconductors is the next stage in the development of spin-wave
devices. The recent progress in synthesis of nanometer YIG films of
high quality on semiconductor substrates \cite{Stog15,Stog15a} and
low relaxation of long-wavelength spin waves in nanometer magnetic
films \cite{Lut12,Lut16} give a possibility to construct spin-wave
devices on semiconductor chips operating in the microwave frequency
band.

Active control and manipulation of spin degrees of freedom in
spin-FET is one of the main problems in
spintronics~\cite{Dat90,Hall03,Schl03,Egues03,Zut04,Sug06,Crow07,Koo09,Chuang15}.
The spin transport in two-dimensional electron gas (2DEG) and large
spin-orbit interaction are essential for realizing spin transport
devices. However, the noneffective spin injection and weak influence
of the gate on electron current flowing through the transistor
channel are known difficulties in spin-FET design. Modulating the
channel conductance by using an electric field to induce spin
precession is performed at low temperatures and has remained elusive
at higher ones~\cite{Koo09,Chuang15}. Furthermore, poor crystal
quality of ferrite films sputtered on GaAs substrates has a
detrimental effect on the device performance \cite{Chen12,Buh95}. At
the same time, it needs to note that YIG films are regarded as
perspective materials in spintronics \cite{SSP13}.

In this paper we describe synthesis of YIG films deposited on GaAs
substrates with AlO${ }_{x}$ layers by ion-beam sputtering and
present results of the characterization of YIG film / AlO${ }_{x}$ /
GaAs heterostructures and their interfaces (see Sec. 2). Magnetic
characteristics of deposited films are deduced from the FMR X-band
spectroscopy (Sec. 3). Spin wave propagation are described in Sec.
4. In Sec. 5 we consider the influence of spin excitations in YIG
films on the electron current flowing through the 2DEG channel
formed at the AlO${ }_{x}$ / GaAs interface. It is found that a high
interaction between spin excitations and 2DEG channel in GaAs-based
substrates can be achieved at the ferromagnetic resonance (FMR)
frequency of YIG films. The inverse effect, the influence of the
electron current on the FMR spectrum, is described in Sec. 6. This
interaction is studied by a detection of $S$-parameters of the
transistor channel at microwave frequencies under the light exposure
and without light illumination. The interaction is enhanced with the
light exposure of the AlO${ }_{x}$ / GaAs interface and with the
microwave power increase. It is found that above the spin-wave
instability threshold the increase of the 2DEG density induced by
light results in essential changes in the FMR spectrum and in the
$S_{21}$-parameter of the channel.

\section{Sample preparation and characterization}

\begin{figure}
\begin{center}
\includegraphics*[scale=0.67]{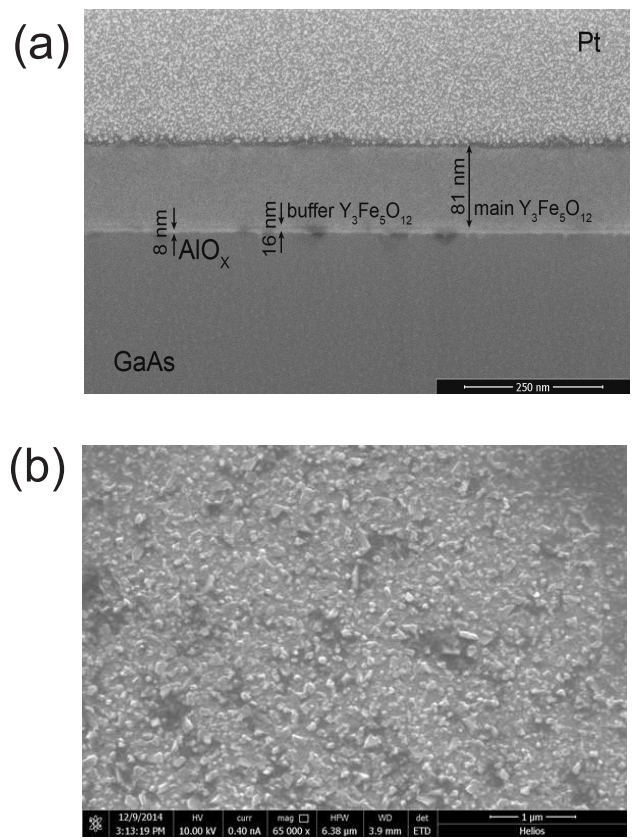}
\end{center}
\caption{(a) Cross-section of the YIG film sputtered on AlO${ }_{x}$
/ GaAs substrate (\# 5). (b) YIG film surface (\# 6). } \label{Fig1}
\end{figure}

YIG films were deposited on GaAs substrates by the two-stage
ion-beam sputtering in Ar + O${ }_{2}$
atmosphere~\cite{Stog92,Nip12}. The $n$-GaAs substrates with
thickness of 0.4~mm had the (100)-orientation. Electrical
resistivity of GaAs chips was measured by the dc four-probe method
at room temperature and was equal to $0.9\times
10^5$~$\Omega\cdot$cm. In order to reduce elastic deformation and
diffusion of Ga ions into the YIG films and to form 2DEG layer in
GaAs substrates, the deposition process was produced on the
amorphouslike nonstoichiometric aluminium oxide layer AlO${ }_{x}$
with the thickness of 8--20~nm ion-beam sputtered previously on
GaAs. The 2DEG layer was formed at the AlO${ }_{x}$ / GaAs
interface~\cite{Koo05,Mann08}. At the first stage a thin (30~nm)
buffer YIG layer was sputtered. After annealing of YIG / AlO${
}_{x}$ / GaAs heterostructure the sputtered buffer YIG layer had a
polycrystal structure. Annealing was performed in the quasi-impulse
regime during 5~min at $590^{\circ}$C in N${ }_{2}$ (samples \# 3,
5, 6, Table \ref{tab1}) and air (samples \# 1, 2, 4) atmospheres
with the pressure of 0.1~Torr. After the deposition and annealing
processes at the first stage, the buffer YIG layer was polished  by
a low-energy (400~eV) oxygen ion beam. The polish procedure
decreased stress tension and dislocations, smoothed areas of
inter-crystallite boundaries and led to reduction of the buffer YIG
layer thickness to 10-16~nm. After this operation the surface of the
buffer layer was suitable to deposit a thicker (main) YIG layer
without lattice mismatch and stress tension. The main YIG layer was
deposited at the second stage. After annealing during 5~min at
$550^{\circ}$C in N${ }_{2}$ (samples \# 3, 5, 6) and air (samples
\# 1, 2, 4) atmospheres with the pressure of 0.1~Torr the sputtered
main YIG layer obtained a polycrystal structure. Cross-section of
the YIG film sputtered on GaAs-based substrate (sample \# 5) is
presented in Fig. \ref{Fig1}a. Cross-section of YIG film of
deposited heterostructure was produced by ion-beam cutting on the
FIB Helios NanoLab 600 Station (FEI Company, USA). YIG film surface
(Fig. \ref{Fig1}b) exposes the roughness caused by large-scale
crystallites of the main YIG layer. The size of crystallites was in
the range of 50--100~nm.

\begin{table}
\caption{Properties of YIG films sputtered on GaAs-based substrates.
\label{tab1}}
\begin{ruledtabular}
\begin{tabular}{ccc|c|cc}
 & &Main layer &Sublayers & \multicolumn{2}{c}{ }\\
$\#$ & Thickness & $4\pi M-H_a$ & $4\pi M-H_a$ & $\Delta H_{\bot}$ &
$\Delta H_{||}$
\\ & $d$ (nm) &  (Oe)& (Oe) & (Oe) & (Oe) \\
\hline
1 & 40 & 1180 & 1464 & 155 & 106 \\
 & & &609 & \multicolumn{2}{c}{ }\\
 & & &-64 & \multicolumn{2}{c}{ }\\
2 & 40 & 1209 & 1372 & 143 & 129 \\
3 & 40 & 990 & 1278 & 245 & 207 \\
 & & &800 & \multicolumn{2}{c}{ }\\
 & & &609 & \multicolumn{2}{c}{ }\\
4 & 250 & 1454 & 1320 & 73 & 155 \\
 & & &1081 & \multicolumn{2}{c}{ }\\
5 & 97 & 651 & & 554 & 651 \\
6 & 964 & 666 & 901 & 175 & 221 \\
 & & &402 & \multicolumn{2}{c}{ }\\
\end{tabular}
\end{ruledtabular}
\end{table}

\begin{figure}
\begin{center}
\includegraphics*[scale=0.7]{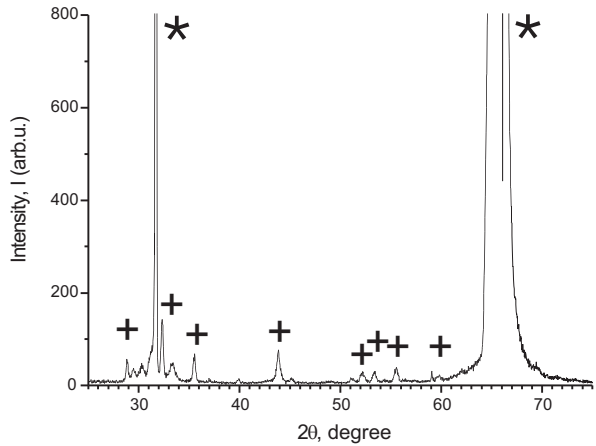}
\end{center}
\caption{XRD spectrum of the sample \# 4. ($*$) marks the substrate
and ($+$) marks the YIG film. } \label{Fig2}
\end{figure}

The structure of YIG films was determined by the X-ray diffraction
(XRD CuK$\alpha$) method and by the energy-dispersive X-ray
spectroscopy. The XRD spectrum confirms the existence of the YIG
phase in the sample (Fig. \ref{Fig2}). It is found that YIG films
are polycrystal and are of homogeneous phase structure. The
spectroscopy methods have shown that the interface layer of GaAs is
enriched with Ga due to the volatility of As ions, however, the
deposited YIG films are not degraded and are not exfoliated from the
GaAs substrates.

\section{Magnetic characteristics of YIG films}

\begin{figure}
\begin{center}
\includegraphics*[scale=0.6]{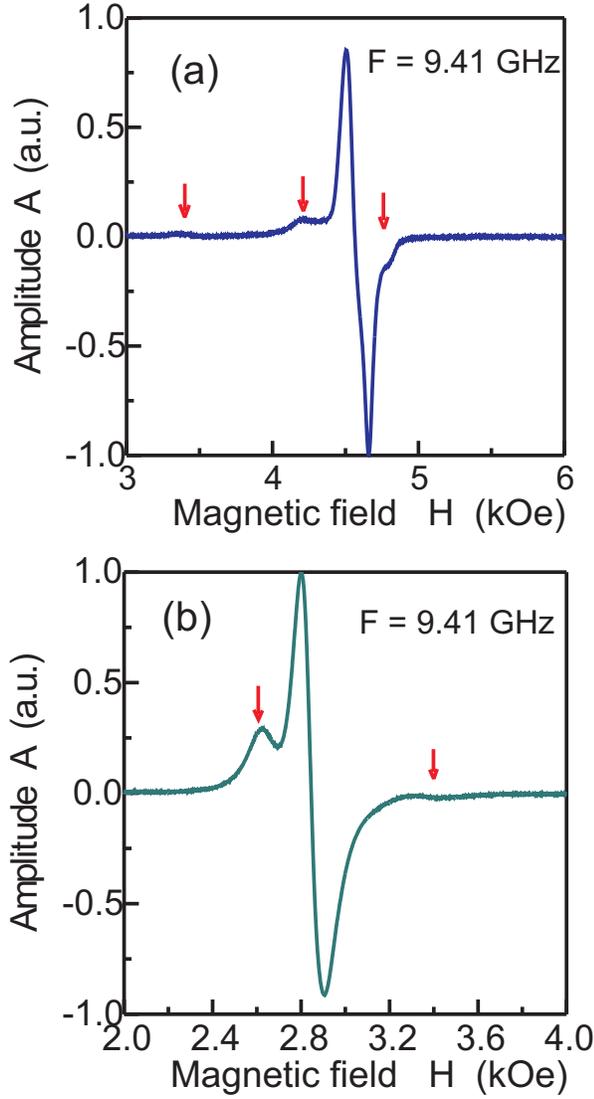}
\end{center}
\caption{FMR spectrum of the YIG film with the thickness of 40 nm
(sample \# 1, Table I) at (a) perpendicular and (b) in-plane
magnetic fields. Arrows correspond to FMR peaks of YIG sublayers. }
\label{Fig3}
\end{figure}

Ferromagnetic resonance of sputtered YIG films was studied by the
X-band electron spin resonance technique. Relating to samples,
applied magnetic field had in-plane and perpendicular orientations.
Using magnetic field sweeping at stabilized frequency $F=9.41$~GHz,
we have read the first derivative of the FMR curve with respect to
the magnetic field $H$. FMR spectrum of the YIG film with the
thickness of 40~nm (sample \# 1, Table \ref{tab1}) at perpendicular
and at in-plane magnetic fields are presented in Fig. \ref{Fig3}.
Arrows correspond to FMR peaks of YIG sublayers.

In order to find magnetic characteristics of the main YIG layer and
sublayers, we use the Lorentzian fitting of experimental curves.
Experimental curves are fitted by the sum of first derivatives of
Lorentzian curves

\begin{equation}
A(H)= \sum^n_i C^{(i)}\frac{\partial L^{(i)}(H)}{ \partial H}
,\label{eq1}
\end{equation}

\noindent where $i$ is the peak number ($i=1$ is the number of the
main YIG layer and $i=2,3,4,\ldots$ are numbers of sublayers),
$C^{(i)}$ is the amplitude,

$$L^{(i)}(H)= \frac{1}{ 1+{(H-H_0^{(i)})^2}/(\Delta H^{(i)}/2)^2}$$

\noindent is the Lorentzian curve, $H_0^{(i)}$ is the peak position,
$\Delta H^{(i)}$ is the FMR linewidth. From the Lorentzian fitting
(\ref{eq1}) we find peak positions $H_0^{(i)}$ and the FMR linewidth
$\Delta H^{(i)}$.

Differences between magnetization and uniaxial anisotropy field
$4\pi M-H_a$ (effective magnetization) of the main YIG layer and YIG
sublayers are found from the FMR peak position $H_{0||}^{(i)}$ of
the corresponding layer at the in-plane magnetic field~\cite{Gur96}

\begin{equation}
F=\gamma\left[H_{0||}^{(i)}(H_{0||}^{(i)}+(4\pi
M-H_a))\right]^{1/2}\label{eq2}
\end{equation}

\noindent and from the FMR peak position $H_{0\bot}^{(i)}$ at the
perpendicular field

\begin{equation}
F=\gamma\left[H_{0\bot}^{(i)}-(4\pi M-H_a)\right],\label{eq3}
\end{equation}

\noindent where $\gamma=$ 2.83~MHz/Oe is the gyromagnetic ratio.
Taking into account Eqs.~(\ref{eq2}) and (\ref{eq3}), values of the
effective magnetization $4\pi M-H_a$ of main YIG layers and YIG
sublayers are found. Effective magnetizations and FMR linewidthes
$\Delta H_{\bot}$ and $\Delta H_{||}$ of main YIG layers are
presented in Table \ref{tab1}. We note that samples \# 1, 2, 4
annealed in the air atmosphere have higher values of the effective
magnetization and lower values of the FMR linewidth than samples \#
3, 5, 6 annealed in the N${ }_{2}$ atmosphere. The reason for the
YIG sublayer formation is not well clear up to now and it is planned
to be clarified in next study.

\section{Spin waves}

\begin{figure}
\begin{center}
\includegraphics*[scale=0.67]{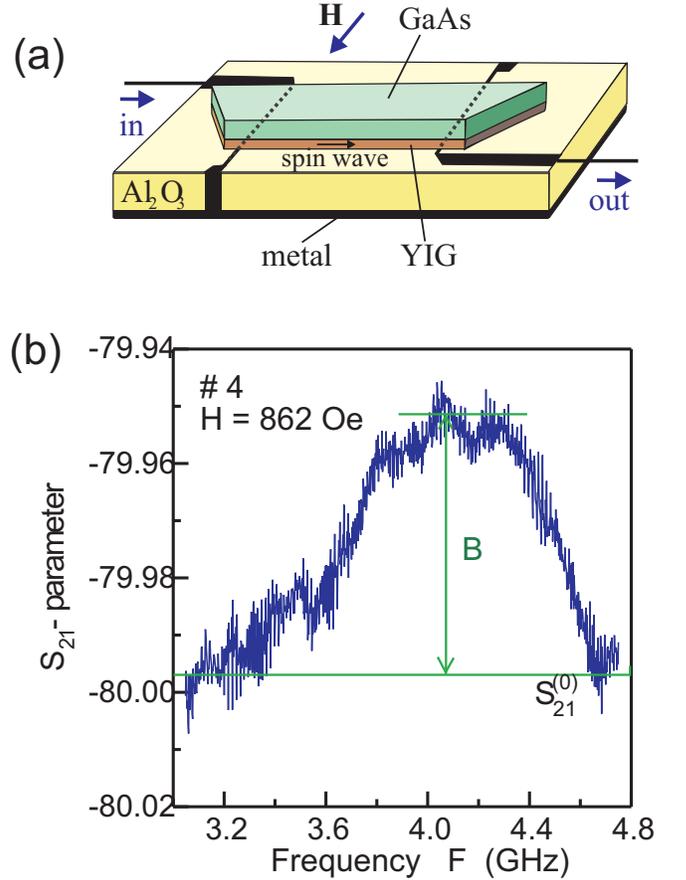}
\end{center}
\caption{ (a) Block diagram of the setup used for spin wave
propagation in YIG / AlO${ }_{x}$ / GaAs structures. (b) The
$S_{21}$-parameter (scalar gain) of spin waves propagated in the
sample \# 4 in the magnetic field $H =$ 862~Oe at the microwave
power $P =$ 7 dBm. } \label{Fig4}
\end{figure}

The FMR linewidth measurement is not sufficient for the
determination of relaxation of spin excitations. The linewidth $\Delta H$ is
formed by relaxation of spin excitations and by magnetic
inhomogeneity of a magnetic film. In order to determine the
relaxation parameters, one should study spin wave propagation
directly. We studied the amplitude-frequency characteristics and the
relaxation of the Damon–-Eshbach surface spin waves~\cite{DE} in the
in-plane oriented magnetic field. The setup is presented in Fig.
\ref{Fig4}a. The spin-wave measurement cell contains microstrip
antennas. The samples are placed on the antenna structure. Antennas
generate and receive spin waves propagated in YIG films. The studied
samples are irregular trapezoidal with sizes of $2\times 6$~mm. The
distance between antennas in the cell was set to 1.2 mm. The
thickness $w$ of antennas is of 30~$\mu$m. The excited spin-wave
wavelength $k$ is given by the thickness $w$ and is in the range
$[0,2\pi /w]$. The antenna length is equal to 2~mm. The measurement setup
contains the Rohde-Schwarz vector network analyzer ZVA-40, which
generates the current flowing in the generating antenna and detects
the current induced by spin waves in the receiving one. We measure
amplitude-frequency characteristics which are the transmission
coefficient $S_{21}$ (the scalar gain) in the frequency range of
3.0--4.8~GHz and in the applied magnetic field $H=$ 862~Oe with the
in-plane orientation. Only for the sample \# 4 we could detect the
transmission coefficient $S_{21}$ (Fig. \ref{Fig4}b). For other
samples, the spin-wave relaxation appeared to be much faster and we
could not detect the spin-wave signals on the receiving antenna.

Measuring the $S_{21}$-parameter, we can estimate the lower bound
$\tau_0$ of the spin-wave relaxation time $\tau$ ($\tau
>\tau_0$)~\cite{Stog15a,Lut16}. For this estimation we
take into account the following approximations.

\noindent (1) We suppose that $|H_a|\ll 4\pi M$.

\noindent (2) In order to calculate spin-wave velocity, we
substitute YIG films with inhomogeneity through thickness by
homogeneous films with higher velocity of propagating spin waves.

\noindent (3) The energy transformations current $\rightarrow$ spin
waves and spin waves $\rightarrow$ current in antenna structure are
perfect and have not losses.

The $S_{21}$-parameter with the voltage induced by spin waves on the
receiving antenna and without spin waves can be written,
respectively, as

$$S_{21}=S_{21}^{(0)}+B=10\lg\frac{(U_s^2+U_0^2)^{1/2}}{U_g}$$
\begin{equation}
S_{21}^{(0)}=10\lg\frac{U_0}{U_g}, \label{eq4}
\end{equation}

\noindent where $U_s$ is the voltage induced by spin waves on the
receiving antenna in the magnetic field $H =$ 862~Oe, $U_0$ is the
voltage on the receiving antenna without a magnetic field and,
consequently, without spin waves, $U_g$ is the voltage on the
generating antenna. We suppose that the spin-wave signals and the
voltage $U_0$ are not correlated. The voltage $U_s$ induced by spin
waves is reduced according to

\begin{equation}
U_s=U_g \exp\left(\frac{-l}{v\tau_0}\right),\label{eq5}
\end{equation}

\noindent where $v$ is the group velocity of spin waves,
$l$ is the distance between antennas. The spin-wave velocity is given
by~\cite{Stan09,Kabos,Gur96,DE}

\begin{equation}
v=\frac{\pi(\gamma 4\pi M)^2d}{2F},\label{eq6}
\end{equation}

\noindent where $F$ is the frequency at the spin wave dispersion
curve at which the wavevector $k\rightarrow 0$, $d$ is the thickness
of the YIG film. Solving the equations (\ref{eq4}), (\ref{eq5}), and
(\ref{eq6}), we find that the spin-wave relaxation time
$\tau_0=39$~$\mu$s and the spin-wave damping parameter, which is
given by~\cite{Gur96,Sparks64}

\begin{equation}
\delta_0=\frac{\Delta\omega_0}{\omega_0}=\frac{1}{2\pi F\tau_0},
\label{eq7}
\end{equation}

\noindent is equal to $1\cdot10^{-3}$. In relation (\ref{eq7})
$\Delta\omega_0=1/\tau_0$ and $\omega_0=2\pi F$. Taking into account
this value of the spin-wave damping parameter, we can find the
contribution of the relaxation process to the FMR linewidth, which
is about 2~\% of the linewidth $\Delta H$. Thus, we can suppose that
the main contribution to the FMR linewidth of the sputtered YIG film
is due to a magnetic inhomogeneity through the film thickness. The
analogous magnetic inhomogeneity has been observed in YIG films
sputtered on GaN substrates~\cite{Stog15a}. The detailed analysis of
the evaluation of the spin-wave damping parameter in inhomogeneous
YIG films is presented in Ref.~\cite{Stog15a}.

\section{Influence of spin excitations on the 2DEG current}

\begin{figure}
\begin{center}
\includegraphics*[scale=0.55]{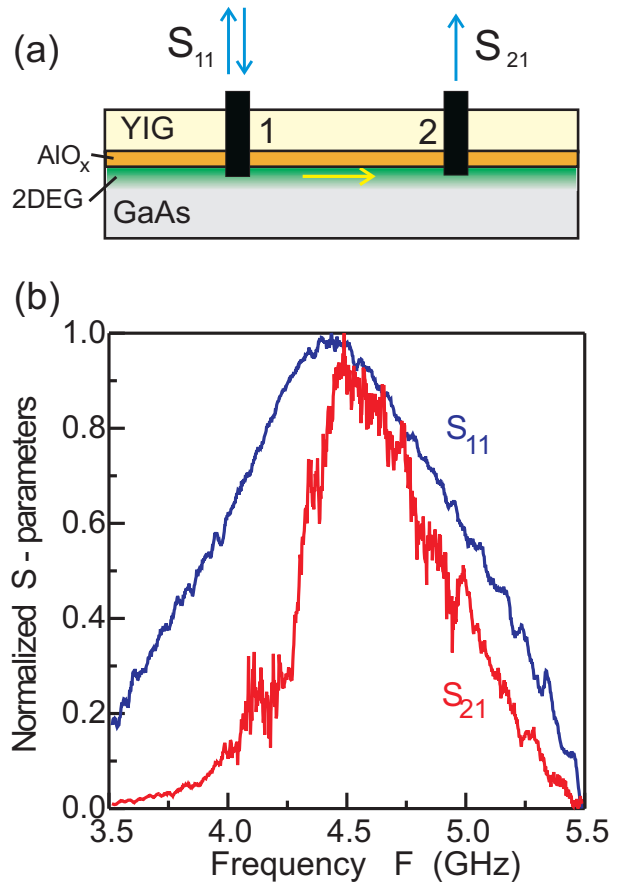}
\end{center}
\caption{(a) Cross-section of YIG / AlO${ }_{x}$ /
GaAs-heterostructure with 2DEG and with contacts 1 and 2. (b)
Normalized $S$-parameters (the transmission coefficient $S_{21}$ and
the voltage reflection coefficient $S_{11}$) measured in the
transistor structure with 2DEG channel formed on the sample \# 3 in
the magnetic field $H=$ 1.107~kOe and at the microwave power $P=$
10~dBm.} \label{Fig5}
\end{figure}

Two-dimensional electron gases are formed at oxide
interfaces~\cite{Koo05,Mann08}. In order to study interaction
between spin excitations in the YIG film and 2DEG in GaAs at the
AlO${ }_{x}$ / GaAs interface, we have performed the transistor
structure with 2DEG channel on the samples \# 3 and \# 6. Electrical
contacts are formed by using the silver paste. We measure
amplitude--frequency characteristics which are the transmission
coefficient $S_{21}$ and the voltage reflection coefficient $S_{11}$
in frequency range of 3.5--5.5~GHz and in applied magnetic fields
$H$ up to 6~kOe with the in-plane orientation. The $S$-parameter
matrix for the 2-port network is defined as
$U_i^{(out)}=S_{ik}U_k^{(in)}$, where $U_i^{(out)}$ is the voltage
wave reflected from the $i$-contact and $U_k^{(in)}$ is the incident
wave at the $k$-contact~\cite{Choma07}. The electrical resistivity
of YIG films is considerably higher than the resistivity of the GaAs
substrate ($0.9\times 10^5$~$\Omega\cdot$cm), consequently, the
channel conductivity between contacts is due to the GaAs 2DEG
interface region (Fig. \ref{Fig5}a). In the FMR frequency band the
YIG-film spin excitations give an influence on the current flowing
through the GaAs channel. The measurement setup contains the
Rohde-Schwarz vector network analyzer ZVA-40, which generates the
current flowing through the 2DEG channel and detects reflected
($S_{11}$) and passed ($S_{21}$) signals. Normalized $S$-parameters
measured in sample \#~3 in the magnetic field $H=$ 1.107~kOe at the
microwave power $P=$ 10~dBm are shown in Fig. \ref{Fig5}b. One can
see that the linewidth of the transmission coefficient $S_{21}$
(593~MHz) is less than the linewidth of the reflection coefficient
$S_{11}$ (1215~MHz). This difference can be explained by a magnetic
inhomogeneity of the YIG film over the thickness $d$. The
$S_{11}$-parameter is formed by the inner volume, upper and YIG /
AlO${ }_{x}$ interface regions of the YIG film near the first
contact. On the contrary, the $S_{21}$-parameter is formed by the
2DEG channel and the neighboring YIG / AlO${ }_{x}$ interface
region. In comparison with inner volume and upper regions, the
neighboring interface region of the YIG film give the greater
contribution to the $S_{21}$-parameter. Since magnetic parameters of
the inner volume, upper and interface YIG regions can be different,
this leads to the difference between $S_{11}$ and $S_{21}$
parameters. The observed interaction between spin excitations and 2D
electrons is of the electromagnetic nature. The alternating magnetic
field of spin excitations induces an alternating electrical field,
which influences on 2D electrons. The observed influence of spin
excitations on the 2DEG current results in modulation of the current
flowing in the 2DEG  channel and, in this sense, one can say that
this modulation is analogous to the action of a gate electrical
potential in FET-structures.

\section{Inverse effect. Influence of the current on spin excitations}

In order to observe the inverse effect -- influence of the current
on spin excitations and to enhance this inverse effect, we have
carried out the experiment under the following conditions: (1) high
channel conductivity, (2) low values of the microwave frequency, at
which the three-magnon decay occurs, and (3) high values of the
microwave power. Increase of the 2DEG current caused by the growth
of the channel conductivity leads to the increase of an alternating
magnetic field acted on the YIG film and at high microwave powers
results in essential changes in the FMR spectrum. According to
\cite{Gur96}, at the three-magnon decay of the FMR excitation at
high microwave powers this influence can be rather high. In the
in-plane magnetic field spin excitations can decay into backward
volume spin waves.

\begin{figure}
\begin{center}
\includegraphics*[scale=0.45]{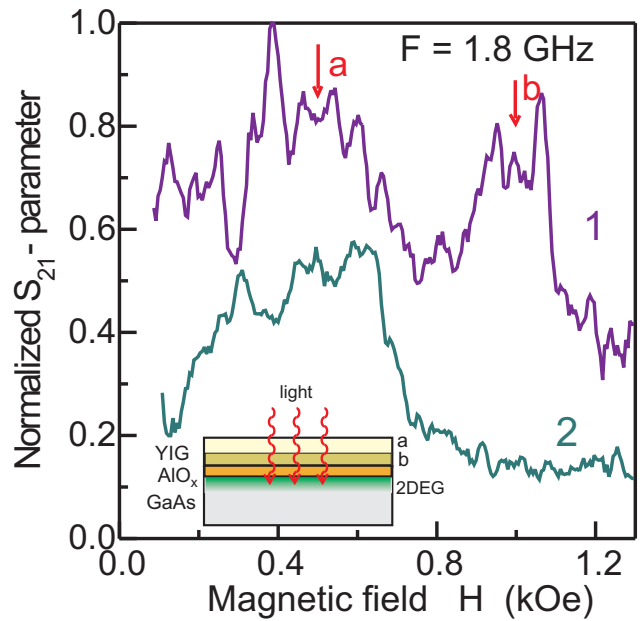}
\end{center}
\caption{The normalized $S_{21}$-parameters measured in the sample \# 6
at the frequency $F=$ 1.8~GHz and at the microwave power $P=$ 14~dBm
(1) under the light exposure and (2) without light. Arrows
correspond to FMR peaks of YIG layers $a$ (without magnon
instability) and $b$ (with magnon instability). } \label{Fig6}
\end{figure}

In order to increase the channel conductivity in the transistor
structure formed on the sample \# 6, the channel was exposed by a
light beam ($\lambda$ = 650~nm, $\varepsilon$ = 1.907~eV) with the
photon energy $\varepsilon$ greater than the GaAs energy band gap of
1.424~eV and less than the YIG band gap of 2.85~eV \cite{Lars75} and
the AlO${ }_{x}$ band gap of 6.5~eV \cite{Nigo12}. The light beam
was linearly polarized with the intensity $W = 81$~mW/cm${ }^2$. The
light exposure leads to electron density increase in the GaAs 2DEG
channel and it is analogous to an action of electric field in FET
structures. Resistance of the channel is reduced from 28.0~M$\Omega$
to 16.9~M$\Omega$. As a result of the light exposure, the local
microwave intensity at neighbouring YIG / AlO${ }_{x}$ interface
region increases. This leads to the three-magnon decay of the FMR
excitation in the YIG interface layer. The magnon instability
process appears at the frequency $F<$ 1.8~GHz and at the microwave
power $P>$ 10~dBm. The normalized $S_{21}$-parameters measured in
the sample \#~6 at the frequency $F=$ 1.8~GHz and at the microwave
power $P=$ 14~dBm under the light exposure and without light are
shown in Fig. \ref{Fig6}. Dependencies are normalized by the maximum
value of the $S_{21}$-parameter measured under the light exposure.
The increase of electrons in the 2DEG channel induced by light leads
to essential changes in the FMR spectrum and in the
$S_{21}$-parameter. One can see that an additional FMR peak $b$
appears in applied magnetic field of 1~kOe. One can observe a
decrease in the height of the peak $a$ and a growth in the amplitude
of the peak $b$, while decreasing frequency of the incident
microwave signal and keeping the microwave power constant and equal
to 14~dBm. Therefore, one can conclude that this leads to an
increase of the thickness of the YIG layer $b$, where the magnon
instability process occurs.

\section{Conclusion}

In summary, we described synthesis of YIG films sputtered on AlO${
}_{x}$ / GaAs substrates, determined their magnetic characteristics,
studied properties of the spin wave propagation and the influence of
spin excitations in YIG films and 2DEG channels formed at the AlO${
}_{x}$ / GaAs interface. It is found that the contribution of the
relaxation process to the ferromagnetic resonance (FMR) linewidth is
about 2~\% of the linewidth $\Delta H$. At the same time, for all
samples FMR linewidths are high. It is supposed that increasing of
the FMR linewidth is due to magnetic inhomogeneity of YIG films.
High interaction between spin excitations and the electron current
flowing through the 2DEG channel formed at the AlO${ }_{x}$ / GaAs
interface is achieved at the FMR frequency of YIG films. On the
other hand, above the spin-wave instability threshold the growth of
the channel conductivity induced by the light illumination results
in essential changes in the FMR spectrum and in the
$S_{21}$-parameter of the channel. The interaction between the spin
excitations in YIG film and 2DEG channel current is increased with
the light exposure of the AlO${ }_{x}$ / GaAs interface and with
microwave power growth. The observed interaction is of great
importance for active control and manipulation of spin degrees of
freedom in field-effect transistors at microwave frequencies.

\section*{Acknowledgments}

This work was supported by the Russian Science Foundation (project
17-12-01314) and the Russian Foundation for Basic Research (project
15-02-06208).

{e-mail: l\_lutsev@mail.ru}

\end{document}